\documentclass{article}
\usepackage{amsfonts}

\newcommand{\beq}{\begin{equation}}
\newcommand{\eeq}{\end{equation}}
\newcommand{\dpl}{\displaystyle}

\begin{document}

\title{On geometry of a special class of solutions to generalised WDVV equations}

\maketitle

\begin{center}

{\bf A.P.Veselov }

\bigskip

{\it Department of Mathematical Sciences, Loughborough University,
Loughborough, Leicestershire, LE 11 3TU, UK
}

\bigskip

{\it Landau Institute for Theoretical Physics, Kosygina 2,

 Moscow, 117940,  Russia

\bigskip

e-mail: A.P.Veselov@lboro.ac.uk,
}

\end{center}

\bigskip

\bigskip

{\small  {\bf Abstract.}
A special class of solutions to the generalised WDVV equations related to 
a finite set of covectors is considered. We describe the geometric conditions 
($\vee$-conditions) on such 
a set which are necessary and sufficient for the corresponding function to satisfy
the generalised WDVV equations. 
These conditions are satisfied for all Coxeter systems but there are also 
other examples discovered in the theory of the generalised Calogero-Moser systems.
As a result some new solutions for the generalized WDVV equations are found.}

\bigskip

\section*{Introduction.}

The WDVV (Witten-Dijgraaf-Verlinde-Verlinde) equations have been introduced first
in topological field theory as some associativity conditions \cite{W,DVV}.
Dubrovin has found an elegant geometric axiomatisation of the these equations 
introducing a notion of Frobenius manifold (see \cite{D}).

What we will discuss in this paper are their generalised versions
appeared in the Seiberg-Witten theory \cite{SW,MMM}
The generalised WDVV equations are
the following overdetermined system of nonlinear partial differential equations:
\beq
F_iF_k^{-1}F_j=F_jF_k^{-1}F_i, \quad i,j,k=1,\ldots,n,
\label{wdvv}
\eeq
where $F_m$ is the $n\times n$ matrix constructed from the third partial
 derivatives of the unknown function $F=F(x^1,\ldots,x^n)$:
\beq
\label{f}
(F_m)_{pq}=\frac{\partial ^3 \, F}{\partial x^m \partial x^p \partial x^q},
\eeq
In this form these equations have been presented by A.Marshakov,A.Mironov and A.Morozov, 
who showed that the Seiberg-Witten prepotential in $N=2$ four-dimensional 
supersymmetric gauge theories satisfies this system \cite{MMM}. 

We will consider the following special class of the solutions to (\ref{wdvv}):
\beq
F^{\mathfrak{A}}=\sum\limits_{\alpha \in \mathfrak{A}} (\alpha,x)^2 \, {\rm log} \, 
(\alpha,x)^2,
\label{imF}
\eeq
where $\mathfrak{A}$ be a finite set of noncollinear vectors $\alpha$ in ${\bf R}^n$.
It is known to be a solution of the generalised WDVV equations in case when $\mathfrak{A}$
is a root system (see \cite{MMM2} for the classical root systems and \cite{MG} for
the general case). In the paper \cite{V} it was observed that the same is true for any
Coxeter configuration and has been found the general geometric conditions on  $\mathfrak{A}$,
which guarantee that (\ref{imF}) satisfies the generalised WDVV equations (the so-called
$\vee$-{\it conditions}). We describe these conditions in the first section below.

It turned out (see \cite{V}) that these conditions are satisfied not only for the root systems
but also for their deformations discovered by O.Chalykh, M.Feigin and the author in the theory 
of the generalised Calogero-Moser systems \cite{ChFV1,ChFV2,ChFV3}. 

The corresponding families of the solutions
 to WDVV equations have the form
\beq
\label{NF1}
F=\sum\limits_{i<j}^{n} (x_i-x_j)^2 \, {\rm log} \, (x_i-x_j)^2 +
\frac{1}{m} \sum\limits_{i=1}^nx_i^2 \,{\rm log} \, x_i^2
\eeq
with an arbitrary real value of the parameter $m$ and
\beq
\label{NF2}
\begin{array}{c}
F= k\sum\limits_{i<j}^{n}\left[ (x_i+x_j)^2 \, {\rm log} \, (x_i+x_j)^2 +
(x_i-x_j)^2  {\rm log} \, (x_i-x_j)^2\right]+ \\
+ \sum\limits_{i=1}^n \left[ (x_i+x_{n+1})^2 {\rm log} \, (x_i+x_{n+1})^2+
 (x_i-x_{n+1})^2 {\rm log} \, (x_i-x_{n+1})^2 \right]+ \\ 
+4m \sum\limits_{i=1}^n x_i^2 \,{\rm log} \, x_i^2+
4l x_{n+1}^2 \,{\rm log} \, x_{n+1}^2,
\end{array}
\eeq
where the real parameters $k, m, l$ satisfy the only relation
\beq
k (2l +1)=2m +1.
\eeq

When $m =1$ the formula (\ref{NF1})  gives the well-known solution to WDVV equations, 
corresponding to the leading perturbative approximation to the exact Seiberg-Witten
prepotential for the gauge group $SU(n+1)$ (see \cite{MMM}).
For the general $m$ it corresponds to the deformation $A_{n}(m)$ of the root
system $A_n$ related to the Lie algebra $su(n+1)$ (see \cite{ChFV1} and below).
Corresponding solution (\ref{NF1}) has been found first in \cite{MMM2} 
(see the formula (3.15) and the calculations after that) 
although the fact that it is related to the configurations 
with non-Coxeter geometry seems to be realised only in \cite{V}.

The second family of the solutions to WDVV equations (\ref{NF2}) has been 
found in \cite{V}.
When $k=m=l=1$ they correspond to the root system
$C_{n+1}$, in the general case - to its deformation  $C_{n+1}(m,l)$
(see \cite{ChFV3} and below). 

In this paper, which is an extended version of \cite{V}, we present also a new 
family of solutions which is related to
a configuration discovered in the theory of integrable Schr\"odinger operators
by Yu.Berest and M.Yakimov \cite{BY}. It has a form $ F (x_1,...,x_k, y_1,...y_l)$,

\beq
\label{NF3}
\begin{array}{c}
F=\sum\limits_{i<j}^{k} (x_i-x_j)^2 \, {\rm log} \, (x_i-x_j)^2 +
 \sum\limits_{p<q}^{l}\mu^{2} (y_p - y_q) ^2 \,{\rm log} \, (y_p - y_q) ^2 +\\ 
+\sum\limits_{i=1}^{k}\sum\limits_{p=1}^{l} (\mu x_i - y_p)^2 \,{\rm log} \,(\mu x_i - y_p)^2,
\end{array}
\eeq
for any integer $k$ and $l$ and arbitrary parameter $\mu$.

The fact that all the families of the configurations discovered so far in the theory
of multidimensional integrable Schr\"odinger operators satisfy the $\vee$-conditions
seems to be remarkable and calls for better understanding.

\section{$\vee$-systems and a particular class of solutions to WDVV equations.}

It is known \cite{MMM,MM}  that WDVV
equations (\ref{wdvv}), (\ref{f}) are equivalent to the equations
\beq
F_iG^{-1}F_j=F_jG^{-1}F_i, \quad i,j=1,\ldots ,n,
\label{fgf}
\eeq
where $G=\sum\limits_{k=1}^n\eta^kF_k$ is any particular invertible linear combination of $F_i$
with the coefficients, which may depend on $x$. Introducing the matrices $\check F_i=
G^{-1}F_i$ one can rewrite (\ref{fgf}) as the commutativity relations 
\beq
\left[ \check F_i, \check F_j \right] =0, \quad i,j=1,\ldots ,n,
\label{com}
\eeq

We will consider the following particular class of the solutions to these equations.

Let $V$ be a real linear vector space of dimension $n$, $V^*$ be its dual space consisting of 
the linear functions on $V$ (covectors),  $\mathfrak{A}$ be a finite
 set of noncollinear covectors $\alpha \in V^*$. 

Consider the following function on $V$:
\beq
F^{\mathfrak{A}}=\sum\limits_{\alpha \in \mathfrak{A}} (\alpha,x)^2 \, {\rm log} \, 
(\alpha,x)^2,
\label{mF}
\eeq
where $(\alpha,x)=\alpha(x)$ is the value of covector $\alpha \in V^*$ on a vector $x\in V$. 
For any basis $e_1,\ldots, e_n$ we have the corresponding coordinates
 $x^1,\ldots , x^n$
in $V$ and the matrices $F_i$ defined according to  (\ref{f}). In a more invariant form
 for any vector $a \in V$ one can define the matrix 
$$
F_a=\sum\limits_{i=1}^n a^iF_i.
$$
By a straightforward calculation one can check that
 $F_a$  is the matrix 
of the following bilinear form on  $V$
$$
F_a^{\mathfrak{A}}=\sum\limits_{\alpha \in \mathfrak{A}} \frac{(\alpha,a)}{(\alpha,x)}
\alpha\otimes \alpha,
$$
where $\alpha\otimes \beta (u,v)=\alpha (u)\beta (v)$ for any $u,v \in V$ and $\alpha, \beta \in V^*$.

Another simple check shows that $G^{\mathfrak{A}}$ which is defined as 
as $F_x^{\mathfrak{A}}$, i.e.
$$
G^{\mathfrak{A}}=\sum\limits_{i=1}^n x^iF_i
$$
is actually the matrix of the bilinear form
\beq
G^{\mathfrak{A}}=\sum\limits_{\alpha \in \mathfrak{A}} \alpha\otimes \alpha,
\label{mG}
\eeq
which does not depend on $x$.

We will assume that the covectors $\alpha \in \mathfrak{A}$ generate $V^*$,
in this case the form $G^{\mathfrak{A}}$ is non-degenerate. 
 This means  that the natural linear mapping
$\varphi_\mathfrak{A} : V\rightarrow V^*$  defined by the formula 
$$
(\varphi_\mathfrak{A}(u),v)=G^{\mathfrak{A}} (u,v), \, u,v \in V
$$
is invertible. We will denote  $\varphi_\mathfrak{A}^{-1}(\alpha),\,  \alpha \in V^*$ as
$\alpha^{\vee}$. By definition 
$$\sum\limits_{\alpha \in \mathfrak{A}} \alpha^{\vee}\otimes \alpha = Id$$
as an operator in $V^*$ or equivalently
\beq
(\alpha,v)=\sum\limits_{\beta \in \mathfrak{A}}(\alpha, \beta^{\vee})(\beta,v).
\label{vee}
\eeq
for any $\alpha \in V^*, v \in V$.
Now according to (\ref{com}) the WDVV equations (\ref{wdvv},\ref{f}) for the function (\ref{mF}) 
can be rewritten as
\beq
\label{ab}
\left[\check F_a^\mathfrak{A}, \check F_b^\mathfrak{A} \right] =0
\eeq
for any $a,b \in V$, where the operators $\check F_a^{\mathfrak{A}}$ are defined as
\beq
\check F_a^{\mathfrak{A}}=\sum\limits_{\alpha \in \mathfrak{A}} \frac{(\alpha,a)}
{(\alpha,x)}\alpha^\vee \otimes \alpha.
\eeq

 A simple calculation  shows that (\ref{ab}) can be rewritten as 
\beq
\label{sc}
\sum\limits_{\alpha \ne \beta, \alpha,\beta \in \mathfrak{A}}
\frac{G^\mathfrak{A} (\alpha^\vee, \beta^\vee)B_{\alpha,\beta}(a,b)}
{(\alpha,x)(\beta,x)}\alpha\wedge \beta \equiv 0,
\eeq
where 
$$
\alpha\wedge \beta=\alpha\otimes \beta-\beta \otimes \alpha
$$ 
and
$$
B_{\alpha,\beta}(a,b)=\alpha\wedge \beta(a,b)=\alpha(a)\beta(b)-\alpha(b)\beta(a).
$$
Thus the WDVV equations for the function (\ref{mF}) are equivalent to the conditions
(\ref{sc}) to be satisfied for any $x,a,b \in V$.

 Notice that WDVV equations (\ref{wdvv},\ref{f}) and, therefore, the conditions (\ref{sc}) 
are obviously satisfied
for any two-dimensional configuration $\mathfrak{A}$. This fact and the structure 
of the relation (\ref{sc}) motivate the following notion of the $\vee$-systems
\cite{V}.

  Remind first that for a pair of  bilinear forms $F$ and $G$ on the vector space $V$
 one can define an eigenvector $e$ as the kernel of the bilinear form $F-\lambda G$
 for a proper $\lambda$:
$$
(F-\lambda G) (v,x)=0
$$
 for any $v \in V$. When $G$ is non-degenerate $e$ is the eigenvector of the corresponding operator 
$\check F=G^{-1} F$:
$$
\check F(e)=G^{-1}F(e)=\lambda e.
$$
Now let $\mathfrak{A}$ be as above any finite set of non-collinear covectors
$\alpha \in V^*$, $G=G^\mathfrak{A}$ be the corresponding bilinear form (\ref{mG}),
which is assumed to be non-degenerate, $\alpha^\vee$ are defined by
(\ref{vee}). Define now for any two-dimensional plane $\Pi \subset V^*$ a form
\beq
G^\mathfrak{A}_\Pi (x,y)= \sum\limits_{\alpha \in \Pi \cap \mathfrak{A}} (\alpha,x)(\alpha,y).
\label{GP}
\eeq

\smallskip

{\bf Definition.} {\it We will say that $\mathfrak{A}$ satisfies the $\vee$-conditions 
if for any plane $\Pi \in V^*$ the vectors $\alpha^\vee, \, \alpha \in \Pi \cap \mathfrak{A}$
are the eigenvectors of the pair of the forms $G^\mathfrak{A}$ and $G^\mathfrak{A}_\Pi$. 
In this case we will call  $\mathfrak{A}$ as $\vee$-system.}

\smallskip

The $\vee$-conditions can be written explicitly as
\beq
\sum\limits_{\beta \in \Pi \cap \mathfrak{A}}
\beta(\alpha^\vee)\beta^\vee=\lambda \alpha^{\vee},
\label{expl}
\eeq
for any $\alpha \in \Pi \cap \mathfrak{A}$ and some $\lambda$, which may depend 
on $\Pi$ and $\alpha$.

If the plane $\Pi$ contains 
no more that one vector from $\mathfrak{A}$ then this condition is obviously satisfied,
so the $\vee$-conditions should be checked only for a finite number of planes $\Pi$. 

If the plane $\Pi$
contains only two covectors $\alpha$ and $\beta$ from $\mathfrak{A}$ then the 
condition (\ref{expl}) means that $\alpha^\vee$ and $\beta^\vee$ are orthogonal with respect 
to the form $G^\mathfrak{A}$:
$$
\beta(\alpha^\vee) = G^\mathfrak{A}( \alpha^\vee,\beta^\vee)=0.
$$

If the plane $\Pi$ contains more that two covectors from $\mathfrak{A}$ this condition
means that $G^\mathfrak{A}$ and $G_\Pi^\mathfrak{A}$ restricted to the plane $\Pi^\vee \subset V$ are
proportional:
\beq
\label{restr}
\left. G_\Pi^\mathfrak{A} \right|_{\Pi^\vee}=\lambda (\Pi) \left. G^\mathfrak{A} \right|_{\Pi^\vee}
\eeq

\smallskip

{\bf Theorem 1.} {\it A function (\ref{mF})
satisfies the WDVV equations (\ref{wdvv}) if and only if the configuration 
$\mathfrak{A}$ is a $\vee$-system }

\smallskip

{\it Proof.}  We have shown above that the function (\ref{mF}) 
satisfies the generalised WDVV equation iff the relations (\ref{sc})
are satisfied. Rewriting these relations as
$$
\sum\limits_{\beta \neq \alpha, \beta \in \mathfrak{A}}
\frac{G^\mathfrak{A} (\alpha^\vee, \beta^\vee) B_{\alpha,\beta}(a,b)}
{(\beta,x)}\alpha \wedge \beta |_{(\alpha,x)=0}\equiv 0
$$
for any $\alpha \in \mathfrak{A}$ it is easy to see that they are equivalent to
$$
\sum\limits_{\beta \neq \alpha, \beta \in \Pi \cap \mathfrak{A}}
\frac{G^\mathfrak{A} (\alpha^\vee, \beta^\vee) B_{\alpha,\beta}(a,b)}
{(\beta,x)}\alpha \wedge \beta |_{(\alpha,x)=0}\equiv 0.
$$
for any $\alpha \in \mathfrak{A}$ and any two-dimensional plane $\Pi$ containing 
$\alpha$ (cf. \cite{ChFV3}).
The last conditions are equivalent to
\beq
\label{pi}
\sum\limits_{\beta \neq \alpha, \beta \in \Pi \cap \mathfrak{A}}
G^\mathfrak{A} (\alpha^\vee, \beta^\vee) B_{\alpha,\beta}(a,b) = 0
\eeq
for any $\alpha \in \mathfrak{A}$ and $\Pi$ such that $\alpha \in \Pi$.

We would like to show that these relations are actually equivalent to the 
$\vee$-conditions. If $\Pi$ contains only two covectors $\alpha$ and $\beta$ 
from $\mathfrak{A}$
then it is obvious since in this case both of these relations are simply saying that
$G^\mathfrak{A} (\alpha^\vee, \beta^\vee) = 0$.

Assume now that $\Pi$ contains more than two covectors. We should show that
in this case $G^\mathfrak{A}$ is proportional $G_{\Pi}^\mathfrak{A}$ after the restriction to $\Pi^{\vee}$.
First of all the relations (\ref{pi}) are obviously satisfied 
if we replace $G^\mathfrak{A}$ by $G_{\Pi}^\mathfrak{A}$:
\beq
\label{pivi}
\sum\limits_{\beta \neq \alpha, \beta \in \Pi \cap \mathfrak{A}}
G_{\Pi}^\mathfrak{A} (\alpha^\vee, \beta^\vee) B_{\alpha,\beta}(a,b) = 0
\eeq
for any $\alpha \in \mathfrak{A}$ and any plane $\Pi$ containing $\alpha$.
This follows for example from the fact that in two dimensions any function
satisfies the generalised WDVV equations, but can be easily checked in a 
straightforward way as well. In particular this immidiately implies that the
$\vee$-conditions are sufficient for (\ref{mF}) to satisfy the generalised WDVV
equations. 

To prove that they are also necessary for this let's suppose that
this is not the case, i.e.  $G^\mathfrak{A}$ is not proportional $G_{\Pi}^\mathfrak{A}$ on $\Pi^{\vee}$.
Then we can find such a constant $c$ that the restriction of
the form $G^\mathfrak{A} - c G_{\Pi}^\mathfrak{A}$ onto $\Pi^{\vee}$ has a rank 1:
$$
G^\mathfrak{A} - cG_{\Pi}^\mathfrak{A}|_{\Pi^{\vee}} = \epsilon \gamma \otimes \gamma |_{\Pi^{\vee}} , \epsilon = +1 or -1.
$$
for some $\gamma \in V^*$.
Without loss of generality we can assume that $(\gamma,\alpha^{\vee}) \geq 0$
for all $\alpha \in \mathfrak{A}$. Let $\alpha_0^{\vee}$ be "the very right"
vector from $\mathfrak{A}^{\vee}$ in the half-plane $(\gamma,v) \geq 0, v \in \Pi^{\vee}$ such that
$B_{\alpha_0,\beta}(a,b)= \alpha_0\wedge \beta(a,b)$ has the same sign for all $\beta \in \mathfrak{A}$.
Now put in the relations (\ref{pi}) and (\ref{pivi}) $\alpha = \alpha_0$ and subtract from
the first relation the second one multiplied by $c$:
$$
\sum\limits_{\beta \neq \alpha_0, \beta \in \Pi \cap \mathfrak{A}}
(G^\mathfrak{A}-c G_{\Pi}^\mathfrak{A}) (\alpha_0^\vee, \beta^\vee) B_{\alpha_0,\beta}(a,b) =
\epsilon (\gamma,\alpha_0^{\vee})\sum\limits_{\beta \neq \alpha_0, \beta \in \Pi \cap \mathfrak{A}}
(\gamma,\beta^{\vee}) B_{\alpha_0,\beta}(a,b) = 0.
$$
Since due to the choice of $\gamma$ and $\alpha_0$ all the summands have the same sign this is possible only if 
all of them are zero,
i.e. $(\gamma,\beta^{\vee})=0$ for all $\beta \in \Pi \cap \mathfrak{A}$ different from $\alpha_0$.
But this contradicts to the assumption that we have more than two noncollinear covectors from $\mathfrak{A}$
belonging to $\Pi$. This completes the proof of the Theorem 1.

{\it Remark.} It is easy to see from (\ref{ab}) that
WDVV equations are equivalent to the commutativity of the
following differential operators of the Knizhnik-Zamolodchikov type
\beq
\label{dkz}
\bigtriangledown_a = \partial_a - \sum\limits_{\alpha \in \mathfrak{A}} 
\frac{(\alpha,a)}{(\alpha,x)}\alpha^\vee \otimes \alpha.
\eeq
As it follows from the Theorem 1 these operators commute and therefore define a flat connection 
on $V$ if and only if the configuration $\mathfrak{A}$ is a $\vee$-system.

\section{Examples of $\vee$-systems and new solutions to generalised WDVV equations.}

Let $V$ be now Euclidean vector  space with a scalar product $(\, ,\,) $, 
and $G$ be any irreducible finite group generated by orthogonal reflections 
with respect to some hyperplanes (Coxeter groups \cite{Burb}).  Let ${\cal R}$ be a set of normal
vectors to the reflection hyperplanes of $G$. We will not fix the length of the normals but
assume that ${\cal R}$ is invariant under the natural action of $G$ and contains exactly two normal 
vectors for any such hyperplane. Let us choose from each such pair of
vectors one of them and form the system ${\cal R}_{+}$:
$$
{\cal R}={\cal R}_{+}\cup ( -{\cal R}_{+}).
$$
Usually  ${\cal R}_{+}$ is chosen simply by taking from ${\cal R}$ vectors which are positive 
with respect to some linear form on $V$. We will call a system ${\cal R}_{+}$ as 
{\it Coxeter system} and  the vectors from ${\cal R}_{+}$ as {\it roots}.

\smallskip

{\bf Theorem 2.} {\it  Any Coxeter  system ${\cal R}_{+}$ is a $\vee$-system.}

\smallskip

 Proof is very simple. First of all the form (\ref{mG})   in this case is 
proportional  to the euclidean structure on $V$ because it is invariant under $G$ 
and $G$ is irreducible. By the same reason this is true for the form $G_\Pi$ (\ref{GP})
if the plane $\Pi$ contains more than two roots from ${\cal R}_{+}$. When $\Pi$
contains only two roots they must be orthogonal and therefore satisfy
 $\vee$-conditions.

 {\bf Corollary.} {\it For any Coxeter system ${\cal R}_{+}$ the function
\beq
\label{Cor}
F=\sum\limits_{\alpha \in {\cal R}_{+}} \,  (\alpha,x)^2  \, {\rm log} \, 
(\alpha,x)^2
\eeq
satisfy WDVV equations (\ref{wdvv}), (\ref{f}).}

When the Coxeter system is a root system of some semisimple Lie algebra this result 
has been proven in \cite{MMM2,MG}. Notice that even when $G$ is a Weyl group
our formula (\ref{Cor}) in general gives more solutions since we have not fixed 
the length of the roots.

\smallskip

It is remarkable that the $\vee$-conditions are also satisfied for the following 
deformations of the root systems discovered in the theory of the generalised Calogero-Moser 
systems in \cite{ChFV1, ChFV2, ChFV3, BY}.

To show this let us make first the following remark. One can consider the class of functions 
related to a formally more general situation when the
 covectors $\alpha$ have also some prescribed multiplicities $\mu_\alpha$
\beq
F^{(\mathfrak{A},\mu)}=\sum\limits_{\alpha \in \mathfrak{A}} \mu_\alpha (\alpha,x)^2 
\, {\rm log} \, (\alpha,x)^2.
\label{frakF}
\eeq
But it is easy to see that this actually will give no new solutions because 
$F^{(\mathfrak{A},\mu)}=F^{\tilde \mathfrak{A}}+$ quadratic terms, where  
$\tilde \mathfrak{A}$ consists of covectors $\sqrt{\mu_\alpha} \alpha$.

  The following configurations $A_n(m)$ and $C_{n+1}(m,l)$ have been introduced
in \cite{ChFV1,ChFV2,ChFV3}. They consist of the following vectors in ${\bf R}^{n+1}$:
$$
A_n(m)=
\left\{
\begin{array}{lll}
e_i - e_j, &  1\le i<j\le n, & {\rm with \,\, multiplicity \,\,}   m,\\
e_i - \sqrt{m}e_{n+1}, &  i=1,\ldots ,n  & {\rm with \,\, multiplicity \,\,}  1,
\end{array}
\right.
$$
and
$$
C_{n+1}(m,l) =
\left\{
\begin{array}{lll}
e_i\pm e_j, &  1\le i<j\le n, &  {\rm with \,\, multiplicity \,\,}   k,\\
2e_i, &  i=1,\ldots ,n  & {\rm with \,\, multiplicity \,\,}   m,\\
e_i\pm \sqrt{k}e_{n+1}, & i=1,\ldots ,n  &  {\rm with \,\, multiplicity \,\,}  1,\\
2\sqrt{k}e_{n+1} & {\rm with \,\, multiplicity \,\,}  l,\\
\end{array}
\right.
$$
where $k = \frac{2m+1}{2l+1}$.

When all the multiplicities are integer the  corresponding  generalisation of
Calogero-Moser system is algebraically integrable, but usual integrability 
holds for any value of multiplicities (see \cite{ChFV1,ChFV2,ChFV3}). 

Notice that when $m=1$ the first configuration coincides with the classical root system
of type $A_n$ and when $k=m=l=1$ the second configuration is the root system of type $C_{n+1}$.
So these families can be considered as the special deformations of these roots systems.

One can easily check that the corresponding sets
$$
\tilde A_n(m)=
\left\{
\begin{array}{ll}
\sqrt{m} \, (e_i - e_j ), &  1\le i<j\le n,\\
\\
e_i - \sqrt{m}e_{n+1}, & i=1,\ldots ,n
\end{array}
\right.
$$
and
$$
\tilde C_{n+1}(m,l) =
\left\{
\begin{array}{ll}
\sqrt{k} e_i\pm \sqrt{k} e_j, &    1\le i<j\le n\\
\\
2\sqrt{m} e_i, & i=1,\ldots ,n\\
\\
e_i\pm \sqrt{k}e_{n+1}, & i=1,\ldots ,n\\
\\
2\sqrt{kl}e_{n+1}, & \\
\end{array}
\right.
$$
with $k=\frac{2m+1}{2l+1}$ satisfy the $\vee$-conditions. To write down the corresponding
solution in a more simple form it is suitable
to make the following linear transformation :
$$
\tilde A_n(m)=
\left\{
\begin{array}{ll}
e_i - e_j , &  1\le i<j\le n,\\
\\
\frac{\dpl 1}{\dpl \sqrt{m}}e_{i}, & i=1,\ldots ,n
\end{array}
\right.
$$
and
$$
\tilde C_{n+1}(m,l) =
\left\{
\begin{array}{ll}
\sqrt{k} (e_i\pm  e_j), & 1\le i<j\le n, \\
\\
2\sqrt{m} e_i, &i=1,\ldots ,n\\
\\
e_i\pm e_{n+1}, & i=1,\ldots ,n\\
\\
2\sqrt{l}e_{n+1}, & \\
\end{array}
\right.
$$
where again $k=\frac{2m+1}{2l+1}$.

Now the corresponding functions $F$ have the form (\ref{NF1}), (\ref{NF2}) written in the Introduction.
\smallskip

The third family of solutions (\ref{NF3}) is related to the configuration which I would
denote as $A_k * A_l(\mu)$

$$
A_k * A_l(\mu) =
\left\{
\begin{array}{ll}
\ e_i - e_j, & 1\le i<j\le k, \\
\\
\ \mu (f_p -  f_q), & 1\le p<q\le l, \\
\\
\ \mu e_i - f_p, &  i=1,\ldots ,k,\, p=1,\ldots ,l \\
\end{array} 
\right.
$$
where $e_i$ and $f_p$ are the basic vectors in ${\bf R}^k$ and  ${\bf R}^l$ correspondingly.
This configuration has been discovered by Yu.Berest and M.Yakimov \cite{BY}
who were looking for a special "isomonodromial deformation" of the Calogero-Moser problem related
to a direct sum of the root systems of types $A_k$ and $A_l$. When the parameter $\mu = 1$
it coincides with the standard $A_{k+l+1}$ root system.
Again the fact that this family of configurations satisfies the $\vee$-conditions
can be checked in a straightforward way.

As a corollary we have the following

 {\bf Theorem 3}. {\it  The functions $F$ given by the formulas (\ref{NF1}), (\ref{NF2}), (\ref{NF3})
 satisfy the generalised WDVV equations.}

It is easy to see that these solutions are really different, i.e. they are not equivalent
under a linear change of variables, which is a symmetry of the generalised WDVV equations. 
Indeed the bilinear form  $G^\mathfrak{A}$ is determined by the configuration  
$\mathfrak{A}$ in an invariant way and thus induces an invariant 
Euclidean structure on $V$ and $V^*$. In particular, all the angles
between the covectors of the configuration are invariant under
any linear transformation applied to the configuration. A simple calculation shows that these
angles depend on the parameters of the families and the only case when these configurations
have the same geometry is when $m=1$ in the first family and $\mu=\pm 1$ in the third one.
In this case we have simply the root system of type $A_N$.

\smallskip

\section{ Concluding remarks.}

At the moment there are no satisfactory explanations why the deformed root systems arisen in the
theory of the generalised Calogero-Moser problems turned out to be $\vee$-systems.
It may be that it is a common geometrical property of all the so-called locus configurations
\cite{ChFV3}. In this connection I'd like to mention that $\vee$-systems can be naturally defined 
in a complex vector space. Their classification seems to be very interesting and important problem.

Another very interesting problem is the investigation of the corresponding almost Frobenius
structures related to these systems (see \cite{D, D2}).
Dubrovin discovered some very interesting duality in the Coxeter case with the Frobenius structures 
on the spaces of orbits of Coxeter groups \cite{D2,D3}. The natural question is whether this can be generalised
for the non-Coxeter $\vee$-systems and what are the corresponding dual Frobenius structures.

\section{ Acknowledgements.}

I am grateful to M.Feigin for the useful comments on the preliminary version of this paper, 
to Yu. Berest for the suggestion to look at the configurations he discovered with M.Yakimov and to B.Dubrovin
for the inspiring discussions at MSRI, Berkeley in February 1999. I am grateful also to A.Marshakov
who attracted my attention to a very important paper \cite{MMM2} where in particular one of the families of 
our solutions (\ref {NF1}) has been first discovered.

\end{document}